\documentclass[aps,prb,twocolumn,amsmath,amssymb,floatfix,longbibliography,article]{revtex4-2}
\usepackage{graphicx}
\usepackage{bm}
\usepackage[colorlinks=true,citecolor=blue]{hyperref}
\usepackage{braket,bbold,color}
\usepackage{tcolorbox}
\usepackage{ifthen}
\usepackage{dsfont}
\usepackage{tikz}
\usepackage{MnSymbol}
\usepackage{graphicx}
\usepackage{placeins}
\usepackage{float}
\usepackage{svg}

\newcommand{\tr}[1]{\mathrm{Tr}\left[ #1 \right]}
\newcommand{\Ham}{\mathcal{H}}
\newcommand{\ko}[1]{^{( #1 )}}

\newcommand{\revision}[1]{{{#1}}}

\begin{document}

\title{Lee-Yang theory of the two-dimensional quantum Ising model}

\author{Pascal M. Vecsei}
\affiliation{Department of Applied Physics, Aalto University, 00076 Aalto, Finland}

\author{Jose L. Lado}
\affiliation{Department of Applied Physics, Aalto University, 00076 Aalto, Finland}
\author{Christian Flindt}
\affiliation{Department of Applied Physics, Aalto University, 00076 Aalto, Finland}

\date{\today}

\begin{abstract}
Determining the phase diagram of interacting quantum many-body systems is an important task for a wide range of problems such as the understanding and design of quantum materials. For classical equilibrium systems,  the Lee-Yang formalism provides a rigorous foundation of phase transitions, and these ideas have also been extended to the quantum realm. Here, we develop a Lee-Yang theory of quantum phase transitions that can include thermal fluctuations caused by a finite temperature, and it thereby provides a link between the classical Lee-Yang formalism and recent theories of phase transitions at zero temperature. Our methodology exploits analytic properties of the moment generating function of the order parameter in systems of finite size, and it can be implemented in combination with tensor-network calculations. Specifically, the onset of a symmetry-broken phase is signaled by the zeros of the moment generating function approaching the origin in the complex plane of a counting field that couples to the order parameter. Moreover, the zeros can be obtained by measuring or calculating the high cumulants of the order parameter. We determine the phase diagram of the two-dimensional quantum Ising model and thereby demonstrate the potential of our method to predict the critical behavior of two-dimensional quantum systems at finite temperatures.
\end{abstract}

\maketitle

\section{Introduction}

Predicting the phase diagram of interacting quantum many-body systems constitutes a central problem in condensed matter and statistical physics~\cite{Sachdev2000,Vojta2003}. Analytically solving a complex many-body problem is rarely possible, and instead numerical techniques have been developed to infer the thermodynamic properties of interacting quantum systems based on their finite-size scaling behavior. For example, powerful methods based on tensor-networks~\cite{PhysRevLett.69.2863,PhysRevB.48.10345,PhysRevLett.75.3537,PhysRevLett.99.220405,Verstraete2008,PhysRevLett.101.250602} and neural networks~\cite{Carleo2017,PhysRevX.7.021021,PhysRevX.8.011006,PhysRevB.97.085104,PhysRevB.95.245134} have recently been used to determine the spin-liquid phase of interacting lattice models~\cite{Yan2011,PhysRevLett.123.207203,PhysRevX.10.021042,PhysRevLett.127.087201,PhysRevB.100.125124,PhysRevX.11.041021} and explore superconductivity in systems of strongly interacting quantum particles~\cite{Jiang2019,PhysRevX.10.031016,PhysRevResearch.2.033073,PhysRevB.100.195141,PhysRevX.11.041013}. Many numerical  approaches rely on evaluating the average value of an order parameter or its Binder cumulant~\cite{Binder1981,Binder1981a,Binder2001,Landau2014}. By contrast, systematic investigations of the fluctuations encoded in the higher cumulants of the order parameter have received less attention.

\begin{figure}[h!]
\includegraphics[width = 0.48\textwidth]{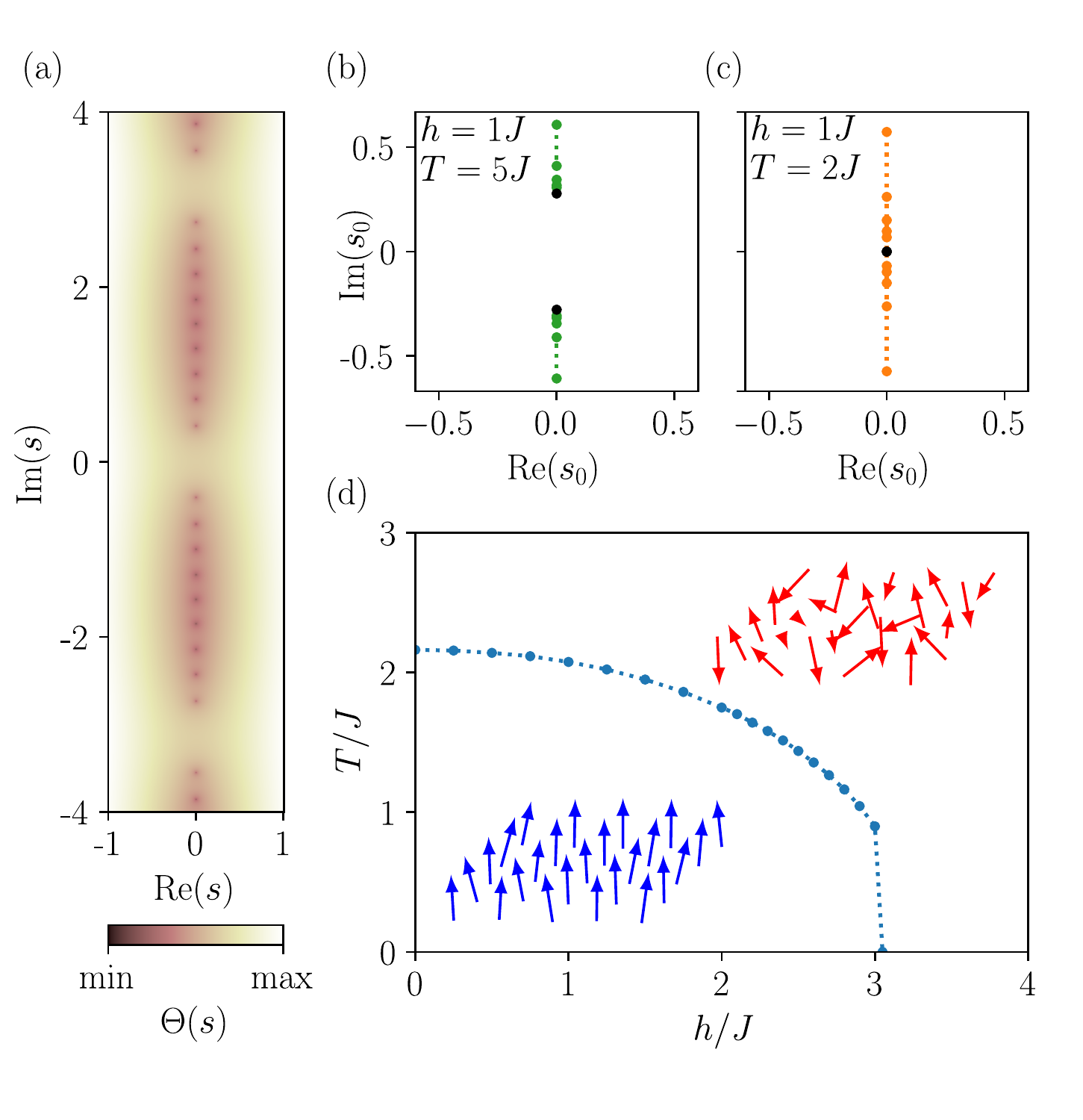}
\caption{\label{introFigure}Phase diagram of the two-dimensional quantum Ising model. (a) Contour plot of the cumulant generating function of the magnetization, $\Theta(s)$ in Eq.~(\ref{eq:CGFdef}) \revision{for $T/J = 5$, $h/J = 1$}. The Lee-Yang zeros appear with dark colors as logarithmic singularities. (b,c) Lee-Yang zeros \revision{closest to the origin of the complex plane,} obtained from the high cumulants of the magnetization for $h/J=1$ and $T/J=5$ in (b) and $T/J=2$ in (c). The black points indicate the convergence points in the thermodynamic limit. (d) Using our cumulant method, we construct the phase boundary between the ordered and disordered phases, illustrated in blue and red, respectively.}
\end{figure}  

For classical equilibrium systems, the theoretical framework of Lee and Yang has been successful in explaining the occurrence of phase transitions~\cite{Lee:1952,Yang:1952,bena2005statistical, blythe2003theleeyang}. Here, the central idea is to consider the zeros of the partition function in the complex plane of the control parameter, for example, the fugacity, an external magnetic field, or the inverse temperature~\cite{fisher1965nature}. For systems of finite size, the partition function is a sum of positive Boltzmann factors, which cannot vanish for real values of the control parameters. However, as the system size is increased, the zeros may approach the value on the real axis, where the system exhibits a phase transition and the free energy becomes nonanalytic in the thermodynamic limit. These ideas have found applications across a wide range of topics, including protein folding in biophysics~\cite{Lee2013,Lee2013a,deger2018leeYang}, Bose-Einstein condensation in atomic physics~\cite{Bormann2000,mulken2001classification,Dijk2015,Gnatenko2017}, and quantum chromodynamics in high-energy physics~\cite{barbour1992complex,ejiri2006leeYang, nagata2015lee,wakayama2019lee,basar2021universality}. Moreover, the Lee-Yang formalism has experienced a surge of interest, as it has been realized that partition function zeros are not only a theoretical concept~\cite{Wei2012,Flindt:2013,Krishnan2019}. They have also been determined in several recent experiments~\cite{Binek1998,peng2015experimental,Brandner2017,francis2021manybody}.

Based on these developments, the ideas of Lee and Yang have been extended to nonequilibrium phase transitions, both in classical~\cite{Arndt2000,Blythe2002} and quantum mechanical systems~\cite{heyl2013dynamical,peotta2020determination}. Recently, we have also taken steps towards a Lee-Yang theory of quantum phase transitions in interacting many-body systems~\cite{kist2021leeYang}, building on a cumulant method that makes it possible to determine the partition function zeros from the fluctuations of the order parameter in small systems~\cite{deger2018leeYang, deger2019determination, deger2020leeYang, deger2020leeYang2}. For classical systems, the moments and cumulants of a thermodynamic observable can be related to the derivatives of the partition function with respect to the control parameter that couples to the observable, for example, energy and inverse temperature or magnetization and magnetic field. Moreover, from the high cumulants, it is possible to find the zeros of the partition function in the complex plane of the control parameter and determine their convergence points in the thermodynamic limit. However, due to noncommuting observables and quantum fluctuations, these relations break down for quantum systems. Therefore, we suggested in Ref.~\cite{kist2021leeYang} instead to consider the zeros of the moment generating function of the order parameter in the complex plane of an auxiliary counting field~\cite{Lamacraft:2008}. As shown for one-dimensional spin lattices at zero temperature, the zeros can then be obtained from the high cumulants of the order parameter, and in the case of a quantum phase transition, the zeros approach the origin of the complex plane in the thermodynamic limit. 

The purpose of the present work is to extend the formalism of Ref.~\cite{kist2021leeYang} to include thermal fluctuations caused by finite temperatures and to demonstrate that the methodology can be applied to two-dimensional quantum many-body systems for which the system size is a well-known limiting factor. Specifically, we consider the two-dimensional transverse-field Ising model, which we, in the present context, refer to as the quantum Ising model to contrast it with its classical counterpart.  Figure~\ref{introFigure}(a) illustrates the positions of the Lee-Yang zeros (in dark colors) for a small lattice, corresponding to singularities of the cumulant generating function, $\Theta(s)$, that we define below. We calculate the high cumulants of the magnetization using tensor-network methods~\cite{Schollwck2011densityMatrix,itensor,hauschild2018finding}, and we can then extract the Lee-Yang zeros that are closest to the real axis as shown in Fig.~\ref{introFigure}(b,c) with increasing system size. Using a finite-size scaling analysis, we can then determine their convergence points in the thermodynamic limit and thereby construct the full phase diagram shown in Fig.~\ref{introFigure}(d). Moreover, within our finite-temperature formalism, we can consider the limit of a vanishing transverse field, and we then recover the partition function zeros for the classical Ising model. As such, our work provides an important connection between the zero-temperature formalism developed in Ref.~\cite{kist2021leeYang} and the classical Lee-Yang theory of phase transitions.

The rest of the paper is organized as follows. In Sec.~\ref{sectionModel}, we introduce the quantum Ising model on a square lattice and briefly discuss its phase diagram. In Sec.~\ref{sec:methods}, we describe the methods used throughout this work, including the finite-temperature Lee-Yang framework that we develop and our tensor-network calculations. In Sec.~\ref{sectionResults}, we present the
results of our work and construct the phase diagram of the quantum Ising model in two dimensions based on our Lee-Yang method. Finally, in Sec.~\ref{sectionConclusion}, we present our conclusions and provide an outlook on possible avenues for further developments. Technical details are deferred to appendices at the end.

\section{Quantum Ising model}
\label{sectionModel}

We focus on phase transitions in the two-dimensional quantum Ising model described by the Hamiltonian
\begin{equation}
\hat\Ham = -J \sum_{\langle i,j \rangle} \hat\sigma_i^z \hat\sigma_j^z - h \sum_i \hat\sigma_i^x,
\end{equation}
where $\hat\sigma_i^{x,z}$ are the Pauli matrices for the spin on site $i$ of the lattice. The coupling between neighboring spins is denoted by $J$, while $h$ is the strength of a transverse magnetic field. The first sum runs over all pairs of neighboring sites, which we indicate by $\langle i,j \rangle$. To minimize edge effects, we impose periodic boundary conditions, while other boundary conditions are discussed in App.~\ref{appendixEffectOfBC}.

The quantum Ising model on a square lattice has a known phase diagram with a quantum phase transition occurring around $h_c\simeq3.04 J$ at zero temperature~\cite{Blote:2002}. Above this critical field, the system is in a paramagnetic phase without spin order in the $z$-direction, and instead the magnetic field forces the spins to point along the $x$-axis. As the magnetic field is decreased and crosses the critical point, the system enters a symmetry-broken phase with a finite magnetization along the $z$-direction. This ordered phase is destroyed as the temperature is increased, and in the absence of a magnetic field, the system undergoes a thermal phase transition at the transition temperature $T_c = 2J/\ln(1 + \sqrt{2})\simeq 2.27J$ with vanishing magnetization above the critical temperature~\cite{schultz1964twoDimensional}, taking $k_B=1$ throughout this work. Figure \ref{introFigure}(d) shows the phase diagram for the quantum Ising model obtained with the Lee-Yang method developed below. As we will see, the phase diagram can be determined from the magnetization cumulants in surprisingly small lattices. Here, we find the phase diagram using numerical calculations of the magnetization and its fluctuations; however, in principle, these may also be measured and thereby allow for experimental predictions of phase diagrams using small lattices of coupled spins.

\section{Methods}
\label{sec:methods}

\subsection{Lee-Yang formalism}
\label{sectionLeeYang}

The starting point of our Lee-Yang formalism is the moment generating function of the order parameter
\begin{equation}
 \chi(s) = \frac{1}{Z} \tr{e^{-\beta \hat\Ham} e^{s \hat M_z } 
 },
 \label{eq:MGF}
\end{equation}
where $Z = \tr{e^{-\beta \hat\Ham}}$ is the canonical partition function, the inverse temperature is denoted by $\beta=1/T$, and we refer to $s$ as the counting field. For the quantum Ising model, the order parameter is the total magnetization
\begin{equation}
 \hat M_z = \sum_i\hat  \sigma_i^z,
\end{equation}
and the moments of the magnetization can be obtained as derivatives with respect to the counting field at $s=0$,
\begin{equation}
\langle  \hat M_z^n  \rangle = \partial_s^n \chi(s)|_{s=0}.
\end{equation}
Moreover, we can define a cumulant generating function,
\begin{equation}
 \Theta(s) = \ln \chi(s),
\label{eq:CGFdef}
\end{equation}
which similarly delivers the cumulants as
\begin{equation}
\llangle  \hat M_z^n  \rrangle = \partial_s^n  \Theta(s) |_{s=0}.
\label{eq:cumudef}
\end{equation}
From these definitions, we obtain the first cumulants in terms of the moments as $\llangle  \hat M_z  \rrangle=\langle  \hat M_z  \rangle$, $\llangle  \hat M_z^2 \rrangle=\langle  \hat M_z^2  \rangle-\langle  \hat M_z  \rangle^2$, $\llangle  \hat M_z^3  \rrangle=\langle  (\hat M_z-\langle  \hat M_z  \rangle)^3\rangle$, and more involved expressions can be derived for the higher cumulants.

In the following, the zeros of the moment generating function in the complex plane of the counting field play the role of partition function zeros in the classical Lee-Yang theory of equilibrium phase transitions~\cite{Yang:1952,Lee:1952,blythe2003theleeyang,bena2005statistical}. In this regard, we observe that $\hat\Ham$ and $\hat M_z$ commute in the absence of a transverse magnetic field ($h=0$), and in that case we can write the moment generating function as
\begin{equation}
    \chi(s) = \frac{1}{Z} \tr{e^{-\beta( \hat\Ham-(s/\beta) \hat M_z )} },
\end{equation}
showing that its zeros become directly related to the partition function zeros for the classical Ising model in the complex plane of a magnetic field in the $z$-direction, $s_k=\beta h_{z,k}$~\cite{deger2020leeYang}. Moreover, at zero temperature, \revision{we write Eq.~(\ref{eq:MGF}) using a complete set of energy eigenstates as
\begin{equation}
    \chi(s) = \frac{\sum_j e^{-\beta E_j} \langle \Psi_j|e^{s \hat M_z }|\Psi_j\rangle}{\sum_j e^{-\beta E_j}}\rightarrow\frac{1}{g}\sum_{i=1}^g\langle \Psi_{0i}|e^{s \hat M_z}|\Psi_{0i}\rangle,
\end{equation}
having taken the limit $\beta E_j\rightarrow\infty$ for each eigenvalue, and
where $\ket{\Psi_{0i}}$ are the orthogonal ground states of the Hamiltonian with degeneracy $g$.}  For the quantum Ising model on a  lattice of finite size \revision{and a non-zero magnetic field}, the ground state is unique ($g=1$), and we denote it by $|\Psi_0\rangle$, such that~\cite{Lamacraft:2008,kist2021leeYang}
\begin{equation}
    \chi(s) = \langle \Psi_0|e^{s \hat M_z}|\Psi_0\rangle.
\end{equation}

The Lee-Yang theory of phase transitions exploits that the partition function for finite-size systems is an entire function that can be expanded in terms of its zeros \cite{Yang:1952,Lee:1952,blythe2003theleeyang,bena2005statistical}. Similarly, the moment generating function allows for a product expansion of the form \cite{conway2012functions,Arfken2012a}
\begin{equation}
 \chi(s) = e^{cs} \prod_k ( 1 - s/s_k),
\end{equation}
where $s_k$ are the complex zeros, and $c$ is a constant that will not be important in the following\revision{, since it only enters the first cumulant and not the higher ones, which are used in our cumulant method.} We can then express the cumulant generating function as 
\begin{equation}
 \Theta(s) = cs+\sum_k\ln ( 1 - s/s_k).
 \label{eq:cgf}
\end{equation}
Moreover, since the moment generating function in Eq.~(\ref{eq:MGF}) is defined as the thermal average of an exponential function, it is always positive for real values of~$s$. Consequently, the zeros cannot be real, and they must come in complex conjugate pairs. In turn, this implies that the cumulant generating function has no singularities along the real axis. We can also introduce the scaled cumulant generating function 
\begin{equation}
    \theta(s) = \lim_{N \to \infty} \Theta(s)/N
\end{equation}
with a well-defined thermodynamic limit for large lattices, where $N=L\times L$ is the total number of sites. The scaled cumulant generating function is similar to the free energy density for equilibrium systems, which is non-analytic for those parameter values, where the system exhibits a phase transition. As shown by Lee and Yang, the singularities occur because the partition function zeros in the thermodynamic limit approach the real axis~\cite{Yang:1952,Lee:1952,blythe2003theleeyang,bena2005statistical}. Similarly, the zeros of the moment generating function may approach the real axis in the thermodynamic limit. However, since the counting field is not a real physical control parameter such as the temperature or the magnetic field, the system only exhibits a phase transition, if the zeros reach $s=0$ in the thermodynamic limit. 

\begin{figure*}
 \includegraphics[width = 1.0\textwidth]{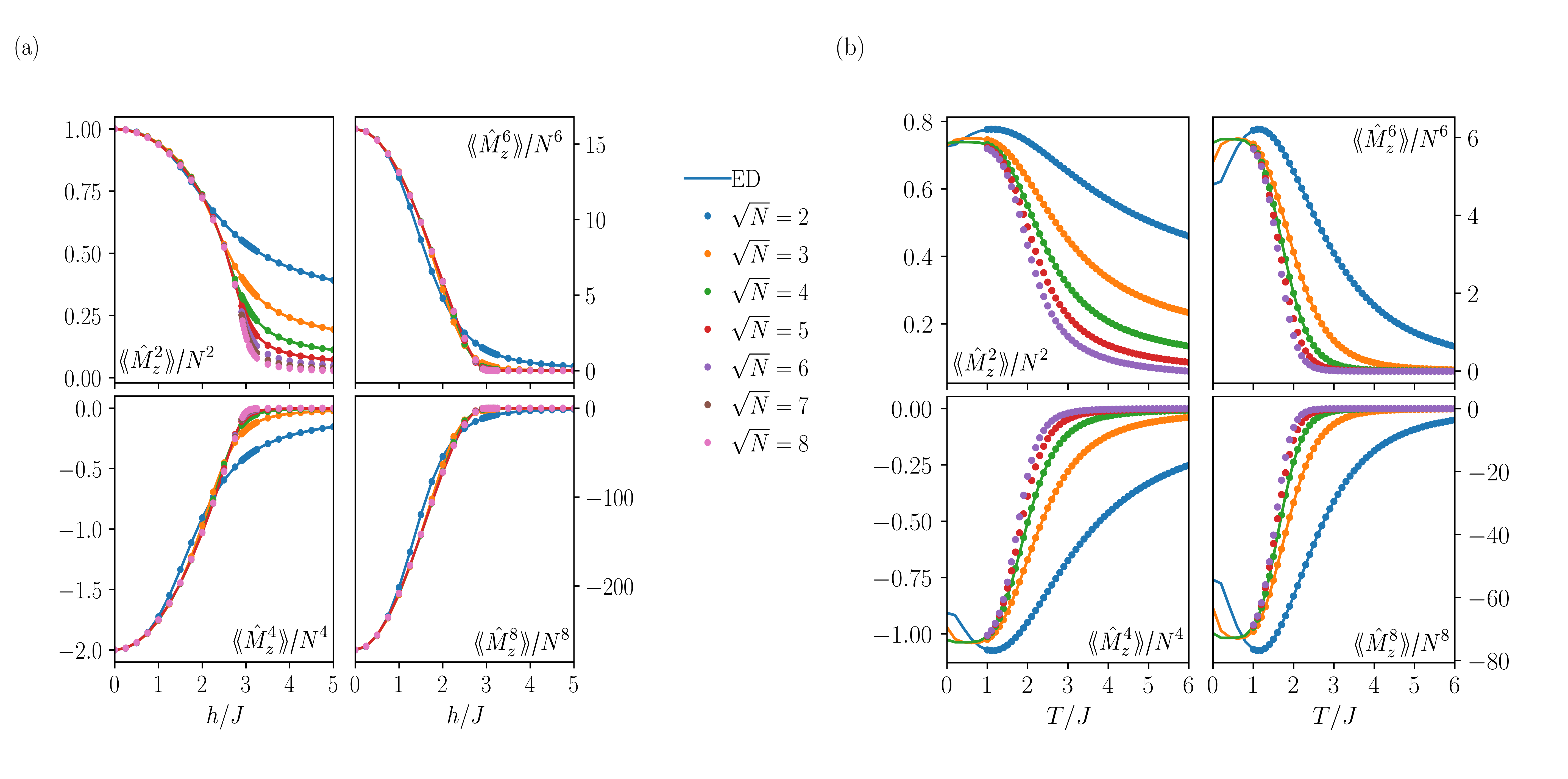}
 \caption{\label{cumulantFigure} Cumulants of the magnetization for different system sizes. (a) Magnetization cumulants calculated at zero temperature as functions of the magnetic field. We compare results obtained with DMRG and exact diagonalization (ED) up to system sizes of $N=L\times L=4\times 4$. (b) Magnetization cumulants as functions of the temperature with the magnetic field $h/J=2$.}
\end{figure*}

Our strategy is now to determine the zeros of the moment generating function from the fluctuations of the magnetization in small lattices. We then extrapolate those results to the thermodynamic limit to determine their convergence points for large systems. To this end, we make use of the cumulant method developed in Refs.~\cite{Flindt:2013,deger2018leeYang,deger2019determination,deger2020leeYang,deger2020leeYang2,kist2021leeYang}, which we briefly describe here for the sake of completeness. First, we use Eqs.~(\ref{eq:cumudef},\ref{eq:cgf}) to express the high cumulants in terms of the zeros as
\begin{equation}
 \llangle \hat M_z^n \rrangle = -  \sum_k \frac{(n-1)!}{s_k^n}\simeq - (n-1)!  \frac{2\cos(n\phi_0)}{|s_0|^n}, n\gg 1 ,
 \label{seriesEq}
\end{equation}
where we have introduced the polar notation $s_k=|s_k|e^{i\phi_k}$ and $s^*_k=|s_k|e^{-i\phi_k}$ for complex conjugate pairs. Also, in the last step, we used that for large cumulant orders the sum is dominated by the pair of zeros, $s_0$ and $s_0^*$, that are closest to $s=0$. The relative contributions from other zeros are suppressed as $|s_0/s_j|^n$, and they can thus be omitted for high enough orders. Importantly, the closest zeros can be obtained from Eq.~(\ref{seriesEq}) and expressed in terms of four consecutive cumulants as~\cite{Flindt:2013,deger2018leeYang,deger2019determination,deger2020leeYang,deger2020leeYang2,kist2021leeYang}
\begin{equation}
  \text{Re}[s_0] \simeq \frac{(n+1) \llangle \hat M_z^n \rrangle \llangle \hat M_z^{n+1} \rrangle - (n-1) \llangle \hat M_z^{n-1} \rrangle \llangle \hat M_z^{n+2} \rrangle }{  2 ( 1 + 1/n) \llangle \hat M_z^{n+1} \rrangle^2 - 2 \llangle \hat M_z^n \rrangle \llangle \hat M_z^{n+2} \rrangle} 
  \label{eq:realpart}
\end{equation}
and
\begin{equation}
  |s_0|^2 \simeq \frac{ n \llangle \hat M_z^n \rrangle^2 -  (n -1) \llangle \hat M_z^{n-1} \rrangle \llangle \hat M_z^{n+1} \rrangle       }{  \llangle \hat M_z^{n+1} \rrangle^2/n -  \llangle \hat M_z^n \rrangle \llangle \hat M_z^{n+2} \rrangle/(n+1)          }. 
  \label{eq:absvalue}
\end{equation}
Thus, by measuring, simulating, or calculating the cumulants of the magnetization, the Lee-Yang zeros can be determined from these expressions. Moreover, if the odd cumulants vanish due to symmetry reasons, we readily find $\text{Re}(s_0)=0$ from Eq.~(\ref{eq:realpart}), and  Eq.~(\ref{eq:absvalue}) simplifies to an expression for the imaginary part reading~\cite{deger2020leeYang}
\begin{equation}
    \text{Im}(s_0)  \simeq \sqrt{2n(2n+1)|\llangle \hat M_z^{2n} \rrangle/\llangle \hat M_z^{2n+2} \rrangle|}. 
  \label{eq:impart}
\end{equation}
In fact, for the quantum Ising model, we easily identify the symmetry, $\hat U^\dag\hat H\hat U= \hat H$, where the unitary operator $\hat U = \prod_i \hat\sigma_i^x$ flips all spins. Inserting identities of the form $\hat U^\dag \hat U=1$ in Eq.~(\ref{eq:MGF}), we find $\chi(s) = \chi(-s)$, since $\hat U^\dag  \hat M_z\hat U= -\hat M_z$, and also $\Theta(s)=\Theta(-s)$, showing that all odd moments and cumulants vanish at any temperature. 

\subsection{Numerical calculations}
\label{sectionNumerics}

To determine the Lee-Yang zeros, we now have to evaluate the high moments and cumulants of the magnetization. Specifically, we need to calculate the moments, 
\begin{equation}
\langle  \hat M_z^n  \rangle = \partial_s^n \chi(s)|_{s=0}= \frac{1}{Z} \tr{\hat M_z^n e^{-\beta \hat\Ham}},
\label{eq:thermalMoments}
\end{equation}
which subsequently are converted into cumulants using the standard recursive relation
\begin{equation}
\langle\!\langle  \hat M_z^n  \rangle\!\rangle =\langle \hat M_z^{n} \rangle-\sum_{m=1}^{n-1}\binom{n-1}{m-1} \langle\!\langle \hat M_z^m \rangle\!\rangle \langle\hat M_z^{n-m} \rangle.
\label{eq:mom2cumu}
\end{equation}
At zero temperature, we find the moments as the ground state averages, $\langle  \hat M_z^n  \rangle = \langle \Psi_0| \hat M_z^n   |\Psi_0\rangle$, which we calculate based on tensor-network methods~\cite{Schollwck2011densityMatrix} with the ground state computed using the density matrix renormalization group (DMRG) method as implemented in ITensor~\cite{itensor}. Although the method was developed for one-dimensional systems with short-range interactions, we can represent the two-dimensional lattice using a snakelike path and find the ground state for lattice sizes up to $L\times L= 8\times 8$ and with bond dimensions up to $\chi_\mathrm{m} = 1000$. 
To ensure numerical stability, the ground state is explicitly symmetrized using $\hat U |\Psi_0\rangle=|\Psi_0\rangle$, and we verify that the cumulants have converged with increasing bond dimension.

\begin{figure*}
 \includegraphics[width = 0.96\textwidth]{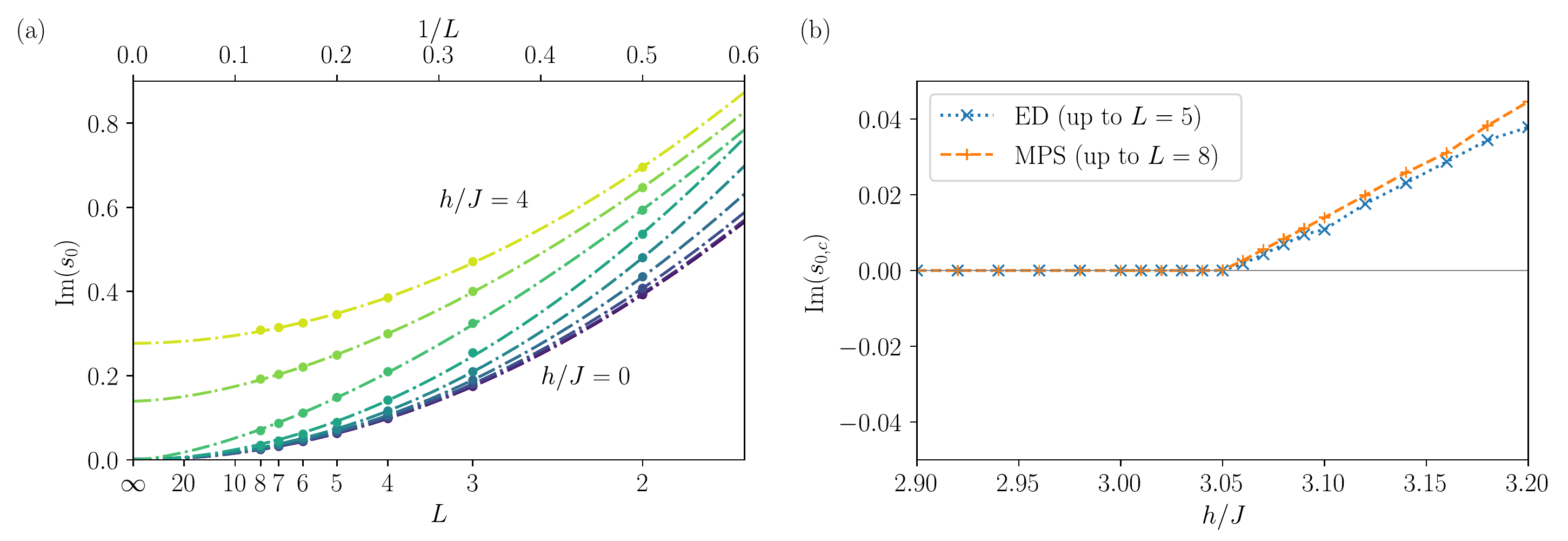}
 \caption{\label{fig:dmrg_t0_extraction_merged} Finite-size scaling of the Lee-Yang zeros. (a) Extrapolation of Lee-Yang zeros to the thermodynamic limit based on Eq.~(\ref{functionFormOfScalingExtrapolation}), which is shown with lines. The results correspond to the magnetic fields $h/J=0.0, 0.5, 1.0, 1.5, 2.0, 2.5, 3.0 ,3.5, 4.0$ and $T=0$.  (b) Extrapolated convergence point as a function of the magnetic field, both from exact diagonalization (ED) and from matrix product states (MPS) calculations, with the former being limit to $L\leq5$. We find a critical field of about $h_{c} \simeq 3.05J$.}
\end{figure*}

Figure~\ref{cumulantFigure}(a) shows zero-temperature cumulants of the magnetization as functions of the magnetic field for several different lattice sizes. For small system sizes, $L\leq 4$, our DMRG calculations are in good agreement with exact diagonalization (ED), providing an important check of our results. We have also ensured that the DMRG calculations have converged with respect to an increased bond dimension, implying that the matrix product state provides a faithful representation of the wavefunction despite the non-local couplings in the lattices, see also App.~\ref{app:num_conv}. For large systems that are away from a phase transition, we expect the cumulant generating function and the cumulants to be linear in the system size, $N=L\times L$. By contrast, in Fig.~\ref{cumulantFigure}(a) we consider small lattices, and as we tune the magnetic field below its critical value of $h_c\simeq3.04 J$, the cumulants scale as $N^n=L^{2n}$, because these values are on the phase transition line between two ferromagnetic states with opposite magnetization.  

To evaluate the moment and cumulants at finite temperatures within the tensor-network formalism, we rewrite the trace operation in Eq.~(\ref{eq:thermalMoments}) as an average with respect to the (unnormalized) pure state 
\begin{equation}
|u_0\rangle= \bigotimes_{i=1}^{L^2}\left(\ket{\uparrow}_i \ket{\uparrow}_{i_\mathrm{aux}} + \ket{\downarrow}_i \ket{\downarrow}_{i_\mathrm{aux}}\right),
\end{equation}
where the spin on each site, $\ket{\sigma}_i$, is now entangled with an auxiliary spin that we denote by $\ket{\sigma}_{i_\mathrm{aux}}$. We can then ``purify'' the trace operation in Eq.~(\ref{eq:thermalMoments}) and write it as  
\begin{equation}
\langle  \hat M_z^n  \rangle = \frac{\langle u_0|e^{-\beta \hat\Ham/2}\hat M_z^n e^{-\beta \hat\Ham/2}|u_0\rangle}{\langle u_0| e^{-\beta \hat\Ham}| u_0\rangle}=\frac{\langle u_{\beta/2}|\hat M_z^n |u_{\beta/2}\rangle}{\langle u_{\beta/2}| u_{\beta/2}\rangle},
\label{eq:purifiedMoments}
\end{equation}
where $|u_{\beta}\rangle=e^{-\beta \hat\Ham}| u_0\rangle$ and $\langle u_{\beta}|=\langle u_{0}|e^{-\beta \hat\Ham}$ are states that have evolved in imaginary time given by the inverse temperature. We have also used that the Hamiltonian and the operator for the total magnetization only act on the physical spins and not the auxiliary ones~\cite{tenpy,binder2015minimally}. The time evolution in imaginary time then follows Ref.~\cite{zaletel2015timeEvolving}. To reduce numerical errors, we explicitly implement powers of the magnetization, $\hat M_z^n$, as a matrix product operator with the bond dimension $n+1$ as described in App.~\ref{appendixExplicitOperator}. In Fig.~\ref{cumulantFigure}(b), we again show the first four even cumulants of the magnetization, however, this time as a function of the temperature with a fixed magnetic field. As we lower the temperature, the required time evolution in imaginary time becomes longer, and our results  become less accurate. Thus, we only show results for $T\geq J$.

\section{Results}
\label{sectionResults}

\begin{figure*}
 \includegraphics[width = 0.96\textwidth]{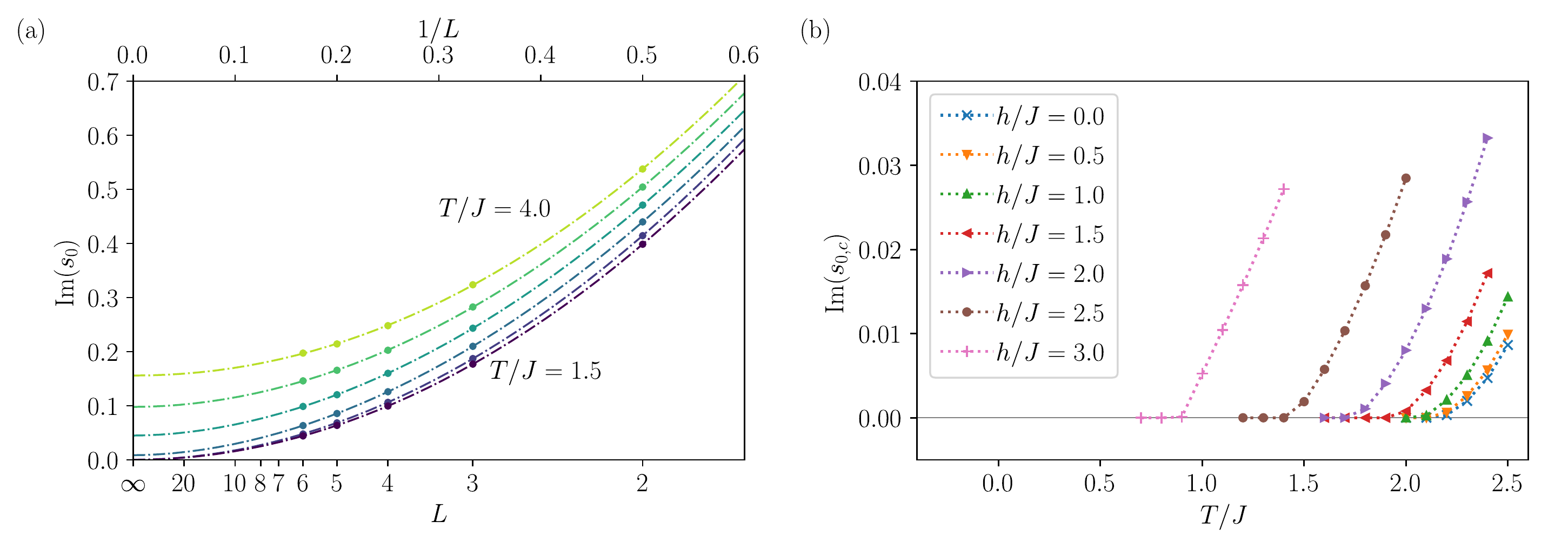}
 \caption{\label{fig:dmrg_tf_extraction}Finite-size scaling of the Lee-Yang zeros at finite temperatures. (a) Extrapolation of the  Lee-Yang zeros to the thermodynamic limit based on Eq. (21), which is shown with lines. We consider the  temperatures  $T/J = 1.5, 2.0, 2.5, 3.0, 3.5, 4.0$ with the magnetic field being $h=0$. (b) For fixed values of the magnetic field, we can determine the corresponding phase transition temperature and thereby construct the phase diagram of the quantum Ising model.}
\end{figure*}

\subsection{Zero temperature}
We are now ready to extract the Lee-Yang zeros from the high cumulants of the magnetization and determine their convergence point in the thermodynamic limit of large system sizes. Specifically, we find the closest pair of Lee-Yang zeros, $s_0$ and $s_0^*$, for different system sizes, $L\times L$, and then extrapolate their convergence points, $s_{0,c}$ and $s_{0,c}^*$, in the limit $L \rightarrow \infty$, which are also known as the Lee-Yang edge singularities. At zero temperature, we determine the Lee-Yang zeros from the magnetization cumulants in the ground state obtained with DMRG. Due to the symmetries of the Hamiltonian, the Lee-Yang zeros are purely imaginary, and in Fig.~\ref{fig:dmrg_t0_extraction_merged}(a) we show the imaginary part of the Lee-Yang zeros as a function of the (inverse) system size.  To determine the convergence points of the Lee-Yang zeros, we assume the scaling form
\begin{equation}
  \text{Im}(s_0)\simeq \text{Im}(s_{0,c})+ \alpha L^{-\gamma},
\label{functionFormOfScalingExtrapolation}
\end{equation}
where $\alpha$ and $\gamma$ are constants. We expect this scaling behavior to hold for systems that are close to criticality, and we have found that it applies for the quantum Ising model in one dimension, where it also works well away from criticality~\cite{kist2021leeYang}. Figure \ref{fig:dmrg_t0_extraction_merged}(a) shows this extrapolation for different values of the magnetic field, and we see that the scaling form fits the available data points well. We can thereby read off the convergence points of the Lee-Yang zeros in the thermodynamic limit. For large magnetic fields, the Lee-Yang zeros remain complex as there is no phase transition. By contrast, at lower fields, the Lee-Yang zeros converge to zero, signaling a quantum phase transition. To determine the critical value of the magnetic field, we show in Fig.~\ref{fig:dmrg_t0_extraction_merged}(b) the convergence points of the Lee-Yang zeros as a function of the magnetic field close to its critical value. \revision{The extrapolated results from ED with maximum sizes of just $L=5$ agree well with  results from DMRG using $L$ up to $8$.} Above the critical point, the Lee-Yang zeros remain complex, and they only vanish as we lower the magnetic field and eventually reach the critical point, which we determine to be $h_{c} \simeq 3.05J$, which is in line with other numerical calculations~\cite{Blote:2002}. \revision{Moreover, at criticality, we expect $\gamma$ to be related to the order parameter $\Delta$ and the dimension~$d$ through the relation $\Delta= d - \gamma$ \cite{kist2021leeYang}.  At criticality, we find $\gamma \simeq 1.46$ and $\Delta \simeq 0.54$ with  $d=2$, which is close to the best known value of $\Delta \approx 0.5181$~\cite{kos2014bootstrapping}.}

\subsection{Finite temperatures}

We can apply the same approach at finite temperatures. Specifically, we find the Lee-Yang zeros from the cumulants of the magnetization for a fixed magnetic field, but with different temperatures. This procedure is illustrated in Fig.~\ref{fig:dmrg_tf_extraction}(a), where we show the Lee-Yang zeros as a function of the system size up to $L=6$. To determine the convergence point in the thermodynamic limit, we again use the interpolation form given by Eq.~(\ref{functionFormOfScalingExtrapolation}), and for the results in Fig.~\ref{fig:dmrg_tf_extraction}(a) with zero magnetic field, we find a critical temperature, which is close to the known expression $T_c = 2J/\ln(1 + \sqrt{2})\simeq 2.27J$. In Fig.~\ref{fig:dmrg_tf_extraction}(b), we have similarly determined the convergence points for different magnetic fields and temperatures. The points, where the curves reach Im$(s_{0,c}) = 0$, then correspond to the phase boundary in the phase diagram. Proceeding in this way, we can map out the phase boundary between the ordered and the disordered phases by considering different temperatures and magnetic fields. 

\subsection{Phase diagram}

Finally, we can  assemble the full phase diagram of the quantum Ising model, which we show in Fig.~\ref{introFigure}(d) based on the results obtained with our cumulant method, both at zero and at finite temperatures. The results are in good agreement with the known phase diagram of the quantum Ising model on a square lattice, both for the critical field at zero temperature and the critical temperature for a vanishing magnetic field. \revision{We note, however, that it is hard to capture the phase boundary at low temperatures close to the critical magnetic field. In that regime, the required time evolution in imaginary time becomes very long, and the results get increasingly inaccurate.} In this context, it is relevant to mention a few issues that require careful consideration when applying our cumulant method. Besides numerical inaccuracies, the determination of the Lee-Yang zeros based on Eq.~(\ref{seriesEq}) requires that high enough cumulant orders are used, so that the sub-leading Lee-Yang zeros can safely be neglected. Thus, one has to check that the results remain unchanged if the cumulant order is increased. Finally, we rely on the scaling ansatz in Eq.~(\ref{functionFormOfScalingExtrapolation}), and one should check it against as large system sizes as possible.

\section{Conclusions}
\label{sectionConclusion}
We have developed a Lee-Yang formalism for quantum phase transitions in interacting quantum many-body systems at finite temperatures. It thereby provides a link between the classical Lee-Yang theory of equilibrium phase transitions and a recent extension to quantum many-body phase transitions at zero temperature. Our methodology considers the zeros of the moment generating function in the complex plane of a counting field that couples to the order parameter. Importantly, in the case in which the system exhibits a phase transition, the zeros move towards the origin of the complex plane as the system size is increased and the thermodynamic limit is approached. The zeros can be determined from the fluctuations of the order parameter, which are encoded in the high cumulants of the order parameter statistics, which in principle are measurable. As a specific application, we have used our Lee-Yang method to construct the phase diagram of the quantum Ising model on a two-dimensional lattice using rather small system sizes combined with a finite-size extrapolation to the thermodynamic limit. To this end, we have employed tensor-network methods to evaluate the high cumulants of the magnetization on Ising lattices of finite size, and we have shown that it is possible to construct the phase diagram of the two-dimensional quantum Ising model. As an outlook on future directions, it may be important to develop methods for evaluating the high cumulants in larger two and ultimately three-dimensional systems, for instance using neural network quantum states. In particular, it may be necessary to treat larger systems, when applying our Lee-Yang method to more complicated spin lattices that could include frustration between neighboring sites.

\acknowledgements
We thank F.~Brange, A.~Deger, and P.~Kumar for useful discussions and acknowledge the computational resources provided by the Aalto Science-IT project as well as the financial support from the Finnish National Agency for Education (Opetushallitus), and from the Academy of Finland through grants (Grants No. 331342, No. 336243, and No. 308515) and through the Finnish Centre of Excellence in Quantum Technology (Projects No. 312057 and No. 312299).

\appendix

\section{Boundary conditions}
\label{appendixEffectOfBC}
\begin{figure}
\includegraphics[width = 0.48\textwidth]{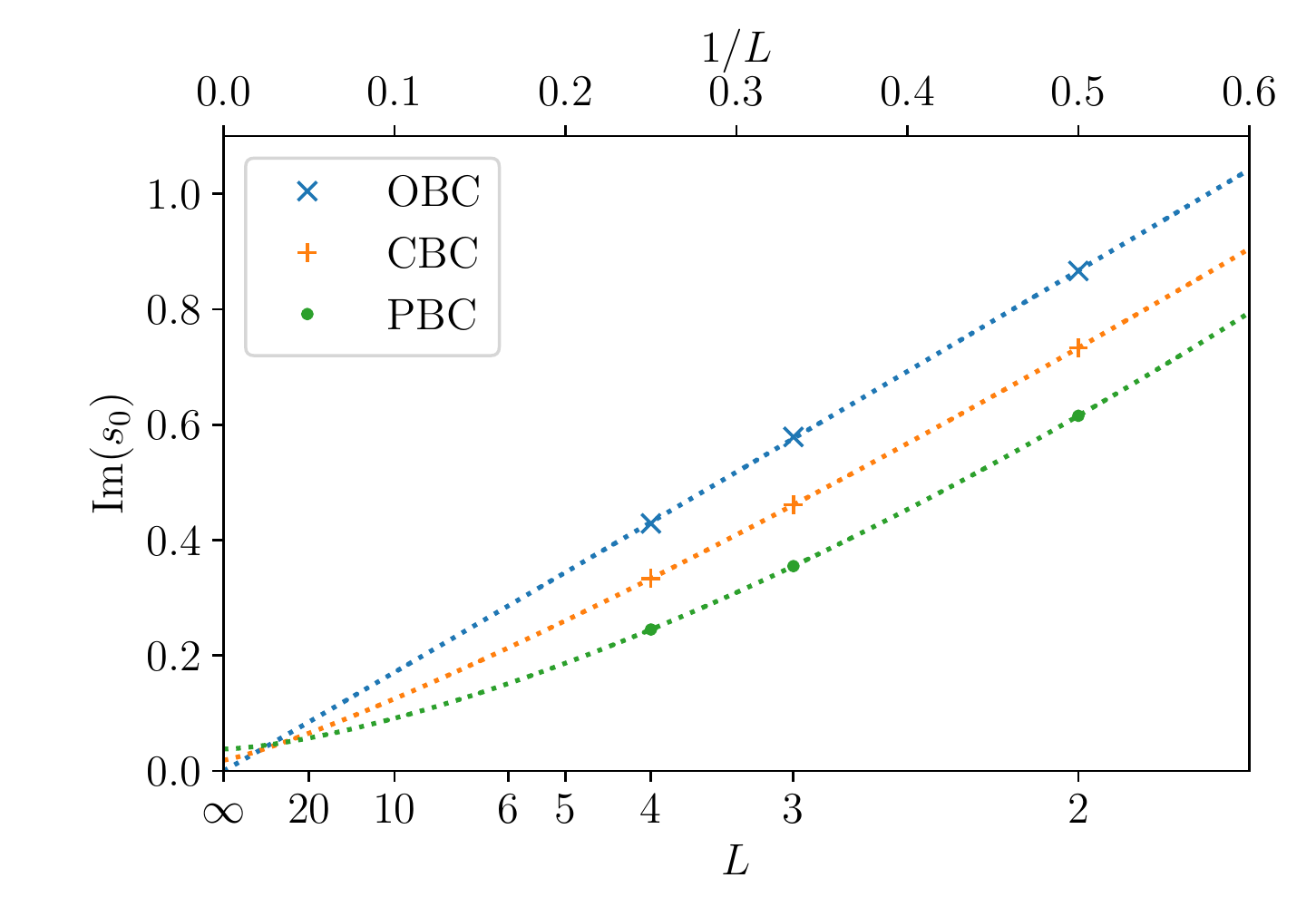}
\caption{\label{fig:comparisonPBC_CBC_OBC_for_DMRG_T0} Boundary conditions. We show Lee-Yang zeros for periodic boundary conditions (PBC), cylindrical (semi-open) boundary conditions (CBC) and open boundary conditions (OBC) for zero temperature and the magnetic field  $h/J=3.2$. The position of the Lee-Yang zeros for finite-size lattices depend on the boundary conditions, while the convergence point in the thermodynamic limit appears to be less sensitive. }
\end{figure}

\begin{figure*}
 \includegraphics[width = 0.96\textwidth]{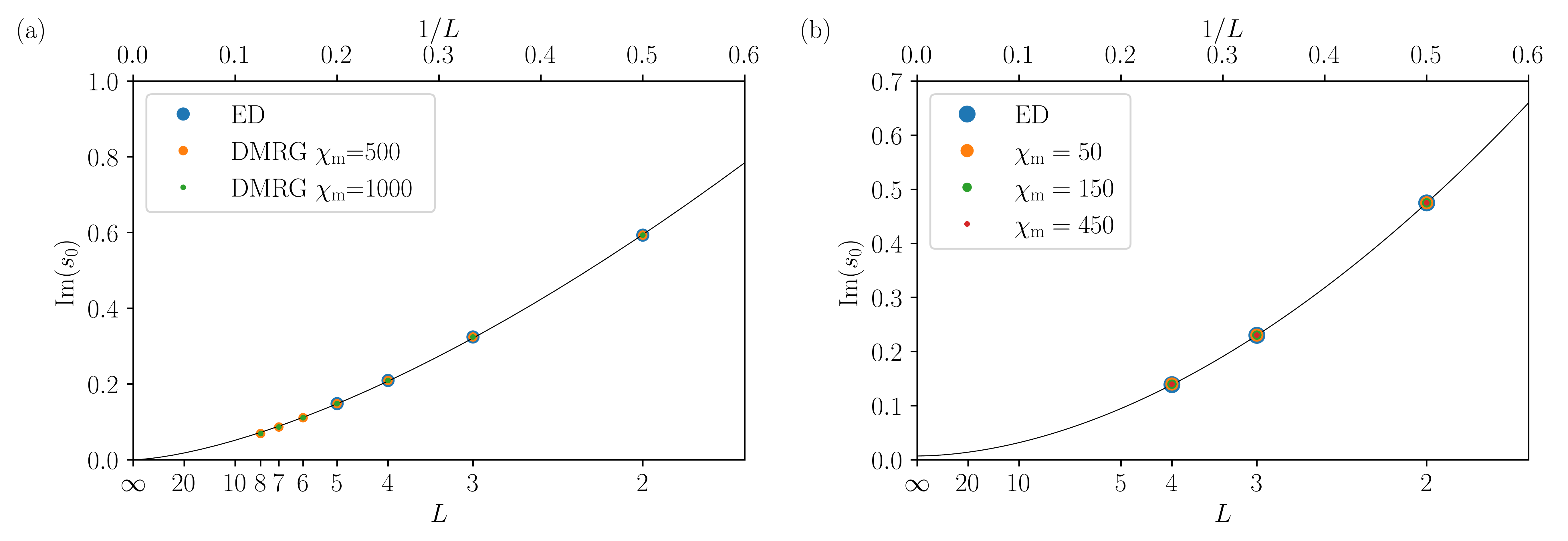}
 \caption{\label{fig:dmrg_t0_comparison_mps_ed} Numerical convergence. (a) Comparison of results based on exact diagonalization (ED) and DMRG with the bond dimension $\chi_\mathrm{m}=500$ and $\chi_\mathrm{m} = 1000$ and the parameter values $T=0$ and $h/J=3.0$. (b) Lee-Yang zeros obtained with ED and results based on matrix product states combined with purification. The parameter values are $h/J=2$ and $T/J = 1$.}
\end{figure*}

Throughout this work, we have imposed periodic boundary conditions to minimize edge effects. It is, however, interesting to investigate the influence of other boundary conditions, which could be important for the small lattices that we consider. Indeed, for a square lattice of size $L\times L$, the number of sites on the boundary is $4L  - 4$, which implies that 75\% of the sites belong to the boundary for a lattice of size $4 \times 4$ and still nearly 50\%  for a lattice of size $8 \times 8$. Therefore, we expect significant effects of the boundary conditions for small lattices, while they should be less important in the thermodynamic limit. In Fig.~\ref{fig:comparisonPBC_CBC_OBC_for_DMRG_T0}, we show Lee-Yang zeros obtained with exact diagonalization for three different boundary conditions: periodic boundary conditions (PBC), cylindrical (semi-open) boundary conditions (CBC) and open boundary conditions (OBC). We see that the position of the Lee-Yang zeros for finite systems depends on the boundary conditions. On the other hand, the extrapolation to the thermodynamic limit is less sensitive to the boundary conditions, and we have used periodic boundary conditions for all calculations in the main text of this paper, because they most closely mimic the conditions in the thermodynamic limit. 

\revision{\section{Details of MPS calculations}
Here, we present details of the MPS calculations used to obtain results at zero and finite  temperatures.

At zero temperature, DMRG was used as implemented in the ITensor library in its Julia version \cite{itensor}. The maximum bond dimension was chosen as $\chi_m = 1000$ and a cutoff of $10^{-9}$ was used. The ground state was converged with 25 sweeps, and the noise level was reduced from $1$ to $10^{-8}$ over the first $13$ sweeps. The 2D lattice was mapped to a 1D chain using a snake-like path, and periodic boundary conditions were imposed. The MPS calculations for the largest systems took less than 1 day on a regular computer node of a local cluster.

For the finite-temperature calculations, the TenPy library was used~\cite{tenpy}. The state was modeled as a purification MPS with two physical indices per site. After initialization of the state at infinite temperature, it was time-evolved using the $W^{II}$ approximation to reach the target temperature \cite{zaletel2015timeEvolving}. The maximum bond dimension was chosen as $\chi_m = 500$, the cutoff was set to $10^{-10}$, and a time-step size of $\Delta \tau = 0.001$ was used. The calculations for the largest systems took about 1 week per value of $h/J$ on one node of a local cluster.
}

\section{Numerical convergence}
\label{app:num_conv}

To check the numerical convergence of the Lee-Yang zeros with increasing bond dimension, we compare our results with exact diagonalization as shown in Fig.~\ref{fig:dmrg_t0_comparison_mps_ed}. In panel (a), we compare results obtained at zero temperature using DMRG and find good agreement with exact diagonalization for lattice sizes up to $L\times L=5\times 5$, which is the upper limit for our numerically exact calculations. A similar agreement is observed in panel (b), where we show results for finite temperatures  obtained by combining matrix product state calculations with purification. 

\section{Explicit expression for $\hat M_z^n$}
\label{appendixExplicitOperator}

For the matrix product calculations, we explicitly implement the operator $\hat M_z^n$ rather than repeatedly applying $\hat M_z$. Any truncation errors appearing during the calculation of the higher moments can be avoided by using this matrix product operator $\hat M_z^n$ with bond dimension $n+1$. The operator is given by
\begin{equation}
 \left(\hat M_z^n\right)_{\sigma_1, ..., \sigma_N, \sigma_1', ..., \sigma_N'} = A\ko 1_{\sigma_1, \sigma_1'} \cdot A\ko 2_{\sigma_2, \sigma_2'} \cdots A\ko N_{\sigma_N, \sigma_N'},
\end{equation}
with the dots denoting matrix multiplication, and we have introduced the matrices $A_{\sigma_i, \sigma_i'}^{(i)} $ with matrix elements
\begin{equation}
    \left(A_{\sigma_i, \sigma_i'}^{(i)}\right)_{m,l} =  \begin{cases}
    0, & l>m \\
    \alpha_m \sigma_{z,\sigma_i, \sigma_i'}^{m-l},  & l=1\\
    \sigma_{z,\sigma_i, \sigma_i'}^{m-l}/(m-l)!, & \text{otherwise}
  \end{cases},
\end{equation}
where we have defined
\begin{equation}
 \alpha_m =  \begin{cases}
    \prod_{k=1}^{n+1-m} (n+1-k) &  m > 1 \text{ and } m \neq N+1\\
    1, & \text{otherwise}
\end{cases},
\end{equation}
and $\sigma_i$ denotes the physical site index. For the first $A$-matrix we take the last row of the matrix, while for the last $A$-matrix just the first column, to obtain an operator in the end. For example, for $n = 6$, we obtain, using the shorthand notation $\sigma$ for $\sigma_{z, \sigma_i, \sigma_i'}$,
\begin{equation}
 A_{ml}^{\sigma_i, \sigma_i'} = \left(
\begin{array}{ccccccc}
 1 & 0 & 0 & 0 & 0 & 0 & 0 \\
 720 \sigma  & 1 & 0 & 0 & 0 & 0 & 0 \\
 360 \sigma ^2 & \sigma  & 1 & 0 & 0 & 0 & 0 \\
 120 \sigma ^3 & \frac{\sigma ^2}{2} & \sigma  & 1 & 0 & 0 & 0 \\
 30 \sigma ^4 & \frac{\sigma ^3}{6} & \frac{\sigma ^2}{2} & \sigma  & 1 & 0 & 0 \\
 6 \sigma ^5 & \frac{\sigma ^4}{24} & \frac{\sigma ^3}{6} & \frac{\sigma ^2}{2} & \sigma  & 1 & 0 \\
 \sigma ^6 & \frac{\sigma ^5}{120} & \frac{\sigma ^4}{24} & \frac{\sigma ^3}{6} & \frac{\sigma ^2}{2} & \sigma  & 1 \\
\end{array}
\right).
\end{equation}

\newpage

%

\end{document}